\documentclass[aps,prb,superscriptaddress,amsmath,amssymb,
showpacs,eqsecnum,twocolumn,floatfix]{revtex4}
\usepackage{graphicx} 
\usepackage{subfigure}
 \usepackage{epsfig,bm} 
\begin{document}
\def\bar{\begin{eqnarray}}
\def\ear{\end{eqnarray}}
\def\beq{\begin{equation}}
\def\eeq{\end{equation}}
\newcommand{\degrees}{\ensuremath{^{\circ}}}
\newcommand{\bsigma}{\mbox{\boldmath$\sigma$}}
\newcommand{\bnabla}{\mbox{\boldmath$\nabla$}}
\title{Induced Triplet Pairing  in clean s-wave Superconductor/Ferromagnet
layered structures}
\author{Klaus Halterman}
\email{klaus.halterman@navy.mil}
\affiliation{Physics and Computational Sciences, Research and Engineering Sciences Department, Naval Air Warfare Center,
China Lake, California 93555}
\author{Oriol T. Valls}
\email{otvalls@umn.edu}
\altaffiliation{Also at Minnesota Supercomputer Institute, University of Minnesota,
Minneapolis, Minnesota 55455}
\author{Paul H. Barsic}
\email{barsic@physics.umn.edu}
\altaffiliation{Current address: Aret\'{e} Associates, 1550 Crystal 
Dr., Arlington,
Virginia 22202}
\affiliation{School of Physics and Astronomy, University of Minnesota, 
Minneapolis, Minnesota 55455}
\date{\today}

\begin{abstract}

We study induced triplet pairing
correlations in clean ferromagnet/superconductor/ferromagnet heterostructures.
The pairing state in the
superconductor is the conventional singlet  s-wave, and 
the angle $\alpha$ between the magnetizations of the two ferromagnetic 
layers is arbitrary.
 We use a numerical fully self-consistent
solution of the microscopic equations and obtain
the time-dependent
triplet correlations via the Heisenberg equations of motion.  
We find that in addition to the usual singlet correlations,
triplet
correlations, odd in time as required
by the Pauli principle, are induced in both 
the ferromagnets and the superconductor.
These time-dependent correlations are largest at times of order
of the inverse of the Debye cutoff frequency, $\omega_D$, and 
we find that within that
time scale they are often spatially very long ranged. We discuss the behavior
of the characteristic penetration lengths that describe these triplet correlations.
We also find that the ferromagnets can locally magnetize
the superconductor near the interface, and that the local
magnetization then undergoes strongly damped
oscillations.
The local density of states exhibits a variety of energy signatures, which
we discuss, as a function of ferromagnetic strength and $\alpha$.

\end{abstract}
\maketitle

\section{Introduction}
Triplet Cooper pairing is no new phenomenon:  
it has long been recognized to be responsible
for superfluidity\cite{sfhe} in $^3{\rm He}$ as well as for superconductivity
in some electronic  materials.  This occurs when the pairing interaction
is in a partial wave with odd $\ell$.
However, recent
observations have raised the possibility of induced triplet
pairing correlations in  s-wave superconductors.
It is a matter of elementary physics  that 
the Cooper pair wavefunction 
must be
antisymmetric under exchange of the two electrons to satisfy the Pauli principle. 
For spatially symmetric s-wave superconductors with decoupled spatial and spin degrees
of freedom, the spin singlet pair is the only possible antisymmetric state of an electron pair.
Triplet pairing states on the other hand, where the
spin state is symmetric,
are 
obviously allowed when the pairing 
is spatially antisymmetric, 
such as in p-wave superconductors. 
Triplet states in systems with
s-wave pairing, even in momentum or coordinate space,
would  naively appear to violate the Pauli principle.  
However, many years ago
Berezinskii proposed\cite{berez} a triplet  state
in superfluid $^3{\rm He}$, which involved spatially symmetric correlations. 
Berezinskii's triplet pairing correlations,
involving different-time pairing, did not
violate the Pauli principle by
virtue of being odd in time, thus 
allowing a triplet state in a system with s-wave interactions.  
While  such a state did not turn out to be  
appropriate to describe superfluidity
in  $^3{\rm He}$, its consideration has led the
way to the study of cases where some sort of time-reversal 
symmetry breaking mechanism
may allow an odd time triplet state to be induced in systems with
spatially symmetric  interactions.

Interest in exotic triplet pairing arises from many quarters.
In the case of two component cold atomic gases\cite{bulgac} with
short-range s-wave interactions, in
which the two species have the same mass but different chemical potentials, it
may be possible to induce a triplet pairing that breaks time reversal symmetry.
In electronic materials, an important issue is the possible existence
of a long range proximity effect in Superconductor/Ferromagnet (SF)
heterostructures. 
Interest in these heterostructures arises in turn
from their possible\cite{zutic} applications. 
Of particular interest
is that the thermodynamic
and transport properties of FSF trilayers are found to depend strongly on the
relative orientation of the magnetization in the two F 
layers.\cite{gu,moraru,bell,visani}  This rather 
well-understood\cite{hv72,dg66,tagirov,buzdin99} fact makes these
structures candidates as spin valves.
  
There have been no unambiguous observations of induced triplet 
correlations in SF structures involving s-wave
superconductors.  However, there have been some enticing
experimental hints in the form of long ranged 
proximity induced superconducting behavior
in SF multilayers with strong exchange fields.
The observed effects are over length scales much larger than those of the
usual
SF proximity effect and more like the 
much longer length scales associated with the 
standard proximity effect between a superconductor and a normal non-magnetic
metal.  
These observations
include measurements in superlattices\cite{pena} with ferromagnetic
spacers and SQUIDs\cite{nelson} with ferromagnetic interlayers.  Superconducting
characteristics, such as a critical temperature and enhanced sub-gap
conductance, have been observed in point contact conductance measurements on
s-wave superconductor/half metallic systems.\cite{kriv}
Perhaps most compelling is the observation of 
a Josephson current through a strong almost half-metallic, 
ferromagnet.\cite{klapwijk06}
All of these experiments indicate long range superconducting
correlations that are not destroyed by a strong exchange field:  
a triplet
state would obviously be consistent with these experiments. 
To fully understand the behavior of these systems, further studies
involving, for example, quantities
sensitive to the gap such as the local density of states
(local DOS), are needed.
Until both better theoretical models and 
more varied experimental observations are made available, 
one cannot conclusively say that there is indeed an induced 
triplet state in these systems,
but at this point the facts fit this
explanation and no better one has been
proposed.\cite{berg05}

Many theoretical studies agree that it is possible to induce this exotic state
in certain SF systems
\cite{berg05,hbv,berg01,berg03,eschrig05a,
fominov07,braude,houzet07,fominov,eschrig05b,buzdin05,asano,yokoyama} 
(perhaps even in the nonmagnetic\cite{tanaka} case) 
with ordinary singlet pairing in S.  Some studies use an ${\rm SFF^{\,\prime}S}$ arrangement in
which the F and ${\rm F^{\,\prime}}$ layers have different magnetization orientations.  Others
assume that a domain structure in a single F layer is responsible for the
symmetry breaking.  Yet others assume an FSF system with different in-plane
magnetization orientations in the F layers.
Whatever the mechanism for the symmetry breaking, such 
arrangements can induce, via proximity effects,
triplet correlations of different kinds. 
Recently it was shown that a Josephson supercurrent 
can exist in a half metal 
by virtue of equal spin triplet pairs and 
spin flip scattering events at the interfaces. \cite{asano} 
To understand and probe the underlying triplet state,
investigations have been done on
conductance spectra in simpler FS structures\cite{linder} with arbitrary magnetization alignment
and with spin active interfaces,
as well as in diffusive\cite{yokoyama} SF junctions, 
through characterizing the possible superconductor symmetry classes consistent 
with Pauli's principle.
With 
the exception of our early work\cite{hbv} on SFS trilayers 
and some recent work on SFS Josephson
junctions,\cite{asano} the studies above
are done in the dirty limit through linearized Usadel-type 
or other  quasiclassical 
equations.  
The disadvantage of a quasiclassical approach is that it is unsuitable for
magnets with an 
exchange field on the order of the Fermi energy.
Thus, it cannot 
properly model a strong ferromagnet, and it does not allow for atomic scale
oscillations in the pair amplitude.  A good quantitative
explanation requires a self-consistent treatment of a fully microscopic model.
Thus, as pointed out in a recent review,\cite{buzdin05} the very
existence of triplet correlations in clean FS structures was until
very recently generally doubted, and these doubts have only very
recently\cite{hbv} been dispelled.

In this paper we 
explore the phenomenon of induced triplet correlations, odd in time,
of clean  FSF structures, where S is
an ordinary s-wave superconductor and  the magnetizations
in the two F layers are rotated by an arbitrary angle $\alpha$. 
We assume strong ferromagnets (up to
the half-metallic limit), and smooth, sharp 
interfaces.  
In this geometry, triplet correlations with total spin projection $m=0$
on the axis of quantization of the Cooper pairs are in general possible and,
when the relative magnetizations (which we assume as usual are both
parallel to the interfaces) are not aligned, triplet components with $m=\pm 1$ are allowed also.
To satisfy the Pauli principle, these spatially symmetric triplet pairing
correlations must be odd in frequency or time.\cite{berez}
That such correlations are allowed, does not mean that they must exist,
nor  that they must exist over an extended spatial range. 
We find, however, 
via a fully self-consistent solution to the microscopic 
Bogoliubov de-Gennes\cite{bdg} (BdG)
equations that such correlations do indeed exist, and that the
penetration depth associated with them can be very long. 
Our use of the BdG equations allows us to study  strong 
ferromagnets. Self-consistency
is fundamental: non-self-consistent solutions are found to
violate the Pauli principle. 
Thus, the  time consuming step of calculating fully
self-consistent solutions is necessary to properly model the proximity effects
which allow for the mixing of superconducting and ferromagnetic orderings that
causes these induced correlations.

In
Sec.~\ref{methods}
of this paper, we discuss the basic  equations  and our method
for numerical self-consistent solution. There, the extraction of the all-important
time dependence via solution of the Heisenberg equations of motion
for the relevant operators is explained in detail. Expressions
for all of the time-dependent triplet correlations are also derived. 
The equations
for the local density of states (local DOS) and the local magnetic moment
(which we use to discuss the reverse proximity effect, that is, the penetration
of the magnetism into the superconductor)  are also presented. In
the next section (Sec.~\ref{results}) we begin by presenting an extensive
discussion of the triplet correlations as a function of position, $\alpha$,
and magnet strength. The appropriate penetration depths are
extracted and discussed. Results for the DOS, the magnetic moment
and the temperature dependence of both triplet and ordinary singlet
correlations are also given. Finally, in Sec.~\ref{conclusions},
a brief conclusion and summary is given.

\section{Methods}
\label{methods}

\begin{figure}
\centering
\includegraphics[width=3.25in]{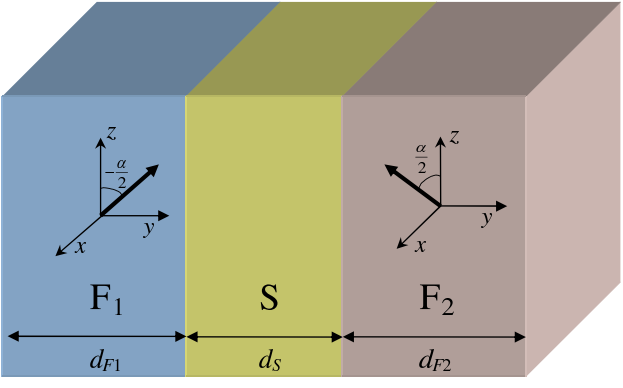}
\caption{(Color~online) Schematic of the FSF junction. The
$y$ axis is normal to the interfaces.
The left ferromagnet layer denoted 
$\rm F_1$ has a  magnetization oriented at an angle $-\alpha/2$ in the $x-z$ plane, while
the other magnet $\rm F_2$, has a magnetization orientation at an
angle $\alpha/2$  in the $x-z$ plane.
All layer widths are labeled.}
\label{fig1} 
\end{figure}

The geometry we consider consists of a planar FSF junction as depicted
in Fig.~\ref{fig1}. The thickness  of the superconducting layer is $d_S$
and the F layers have thicknesses $d_{F_{1}}$ 
and $d_{F_{2}}$. The system is assumed to  be infinite 
in the plane perpendicular to the layers, which we label as our $x-z$
plane. The magnetizations of the F layers, which are in this plane, form an
angle $\pm \alpha/2$ with the $z$ axis, which is 
that of the direction of quantization
of the spins. 

Our starting point is the Bogoliubov de-Gennes (BdG) equations\cite{bdg} for the
system under consideration.
The derivation of the BdG equations for
the case of interest requires some care  with the conventions for all operator
phase factors, which are not universally agreed upon
in the literature, and which may
give rise to
different signs in some of the  equations  below.
We write the effective BCS Hamiltonian, ${\cal H}_{\rm eff}$,
as 
\begin{widetext}
\begin{equation}
{\cal H}_{\rm eff} =\int d^3 r\Bigl\lbrace 
\sum_\alpha \psi^\dagger_\alpha({\bf r})  {\cal H}_{\rm e}  
\psi_\alpha({\bf r})
 + 
\frac{1}{2} [\sum_{\alpha,\beta}  (i\sigma_y)_{\alpha \beta} \Delta({\bf r}) \psi^\dagger_\alpha({\bf r}) \psi^\dagger_\beta({\bf r})+ \rm{h.c.}] -\sum_{\alpha,\beta} \psi^\dagger_\alpha({\bf r})({\bf h} \cdot \bsigma)_{\alpha \beta}\,\psi_\beta({\bf r}) \Bigr\rbrace,
\end{equation}
\end{widetext}
where ${\cal H}_{\rm e}=-1/(2m) \bnabla^2-E_F+U({\bf r})$,
\mbox{\boldmath $\sigma$} are the set of Pauli matrices, spin is denoted by
Greek indices, and
as usual, we represent the magnetism of
the F layers by an effective exchange Stoner energy ${\bf h(r)}$ which will 
in general
have components in both the transverse ($x,z)$ directions. The
spin independent scattering potential is denoted
$U({\bf r})$, 
and $ \Delta({\bf r})$ is the usual pair potential.

To diagonalize 
the effective Hamiltonian, the field operators $\psi^\dagger_\alpha$ and
$\psi_\alpha$ are expanded by means of a Bogoliubov transformation,
which, for our phase convention, we write\cite{song} as: 
\begin{subequations}
\label{bv}
\begin{align}
\psi_{\uparrow}({\bf r})&=\sum_n \left(u_{n\uparrow}({\bf r})\gamma_n - v_{n\uparrow}({\bf r})\gamma_n^\dagger\right), \\ 
\psi_{\downarrow}({\bf r})&=\sum_n \left(u_{n\downarrow}({\bf r})\gamma_n + v_{n\downarrow}({\bf r})\gamma_n^\dagger\right),
\end{align}
\end{subequations}
where $u_{n\alpha}$ and $v_{n\alpha}$ are the quasiparticle 
and quasihole amplitudes,
 and 
$\gamma_n$ and $\gamma_n^\dagger$ are the
Bogoliubov quasiparticle annihilation and creation operators, respectively.

We require that the transformations in Eqs.~(\ref{bv}) diagonalize ${\cal H}_{\rm eff}$,
\begin{subequations}
\label{comm2}
\begin{align}
[{\cal H}_{\rm eff},\gamma_n]&=-\epsilon_n\gamma_n, \\
[{\cal H}_{\rm eff},\gamma^\dagger_n]&=\epsilon_n\gamma^\dagger_n.
\end{align}
\end{subequations}
One can also take the commutator $[\psi_{\alpha}({\bf r}),{\cal H}_{\rm eff}]$. 
With the magnetizations in the $x-z$ plane as explained above, this
gives the following,
\begin{widetext}
\begin{subequations}
\label{comm}
\begin{align}
[\psi_{\uparrow}({\bf r}),{\cal H}_{\rm eff}]&=({\cal H}_{\rm e}-h_z)\psi_{\uparrow}({\bf r})-h_x\psi_{\downarrow}({\bf r})+ 
\Delta({\bf r})\psi^\dagger_{\downarrow}({\bf r}), \\
[\psi_{\downarrow}({\bf r}),{\cal H}_{\rm eff}]&=({\cal H}_{\rm e}+h_z)\psi_{\downarrow}({\bf r})-h_x\psi_{\uparrow}({\bf r})- 
\Delta({\bf r})\psi^\dagger_{\uparrow}({\bf r}).
\end{align}
\end{subequations}
Inserting (\ref{bv}) into (\ref{comm}) and using Eqs.~(\ref{comm2})
yields the general spin-dependent BdG equations,
\begin{align}
\begin{pmatrix} 
{\cal H}_0 -h_z(y)&-h_x(y)&0&\Delta(y) \\
-h_x(y)&{\cal H}_0 +h_z(y)&\Delta(y)&0 \\
0&\Delta(y)&-({\cal H}_0 -h_z(y))&-h_x(y) \\
\Delta(y)&0&-h_x(y)&-({\cal H}_0+h_z(y)) \\
\end{pmatrix}
\begin{pmatrix}
u_{n\uparrow}(y)\\u_{n\downarrow}(y)\\v_{n\uparrow}(y)\\v_{n\downarrow}(y)
\end{pmatrix}
=\epsilon_n
\begin{pmatrix}
u_{n\uparrow}(y)\\u_{n\downarrow}(y)\\v_{n\uparrow}(y)\\v_{n\downarrow}(y)
\end{pmatrix},\label{bogo}
\end{align}
\end{widetext}
where
the single particle Hamiltonian ${\cal H}_0$ is defined as,
\begin{equation}
{\cal H}_0\equiv\frac{\widehat{p}^{\,2}_{y}}{2m}+\varepsilon_{\perp} 
-E_F + U(y).
\end{equation}
 A plane wave factor $e^{i{\bf k_\perp \cdot r}}$ 
has been canceled in both sides of Eq.~(\ref{bogo}).
The longitudinal momentum operator, $\widehat{p}_{y}$, is given by,
$\widehat{p}_{y}=-i \partial/\partial y$, $\varepsilon_{\perp}$ is  the kinetic 
energy of the transverse modes,  $\Delta(y)$ is the self-consistent pair 
potential, and $U(y)$
is a scalar potential representing interface scattering characterized by
a delta function of strength $H_B$.
The ferromagnetic exchange 
field ${\bf h}(y)=(h_x(y),0,h_z(y))$,
vanishes in the S layers. 
We have $h_x(y)=h_0 \sin(-\alpha/2)$ and $h_z(y)=h_0 \cos(-\alpha/2)$ in 
the $\rm F_1$ layer, where
$h_0$ is the magnitude of the exchange field, while in $\rm F_2$,
$h_x(y)=h_0 \sin(\alpha/2)$, and $h_z(y)=h_0 \cos(\alpha/2)$.
We refer to Fig.~\ref{fig1} for details.
The dimensionless parameter $I\equiv h_0/E_F$ conveniently
characterizes the strength of the magnetism. One thus has
$I=1$ in the half metallic limit.
If we take $\alpha=0$ or $\pi$, i.e.
the magnetizations of both layers lie along the same direction,
or if there is only one F layer, 
then $h_x=0$ and we recover the simpler form of the BdG 
equations used\cite{hv69,hv72} in other contexts. 
We can find the quasiparticle amplitudes on a different quantization axis in
the $x-z$ plane forming an angle $\alpha^\prime$ 
with $z$, by performing a spin rotation 
$\Phi_n \rightarrow \widehat{U}(\alpha^\prime)\Phi_n$ with, 
\begin{equation}
\widehat{U}(\alpha^\prime)=
\cos(\alpha^\prime/2)\hat{1}\otimes\hat{1} - i\sin(\alpha^\prime/2) \rho_z
\otimes \sigma_y,
\label{rot}
\end{equation}
where we have introduced \mbox{\boldmath $\rho$} as a set of Pauli-like matrices in 
particle-hole space.
It is convenient to use the matrices ${\bm \rho}$ 
in conjunction with the
ordinary Pauli matrices  ${\bm \sigma}$ in spin space 
to rewrite Eqs.~(\ref{bogo})
in the more compact but perhaps less transparent way:
\begin{equation}
\left[
\rho_z\otimes\left({\cal H}_0 \hat{\bf 1} -h_z\sigma_z\right)+\left(\Delta(y)\rho_x
-h_x \hat{\bf 1}\right)\otimes \sigma_x \right]{\Phi}_n=\epsilon_n{\Phi}_n,
\label{bogoshort}
\end{equation}
where 
${\Phi}_n
\equiv(u_{n\uparrow}(y),u_{n\downarrow}(y),v_{n\uparrow}(y),v_{n\downarrow}(y))^{\rm T}$, 
with the superindex denoting transposition.

The usual self consistency condition
relates the spectrum
obtained from Eq.~(\ref{bogo}) to the inhomogeneous pair potential $\Delta(y)$ by an appropriate sum over states:
\begin{equation}  
\label{del2} 
\Delta(y) = \frac{g(y)}{2}{\sum_{n}}^\prime
\left[u_n^\uparrow(y)v^\downarrow_n (y)+
u_n^\downarrow(y)v^\uparrow_n (y)\right]\tanh(\epsilon_n/2T), 
\end{equation} 
where the prime on the sum indicates that only those positive energy states 
with energy less than 
the pairing interaction energy cutoff, 
$\omega_D$, are included, and $T$ is the temperature.
 The function $g(y)$
vanishes in the F layers while in the S layers
it takes the value of the usual BCS 
{\it singlet} coupling constant in the S material.

With an appropriate choice of basis,\cite{hv69,hv70} 
Eqs.~(\ref{bogo})
can be cast into a finite $4N \times 4N$ dimensional matrix eigenvalue 
system. In dimensionless form, it reads,
\begin{widetext}
\begin{equation}
\label{mbogo}
\begin{bmatrix} 
H_0-H_z&-H_x&0&D \\
-H_x&H_0+H_z&D&0 \\
0&D&-(H_0-H_z)&-H_x \\
D&0&-H_x&-(H_0+H_z)
\end{bmatrix}
\Psi_n
=
\widetilde{\epsilon}_n
\,\Psi_n,
\end{equation}
where 
$\widetilde{\epsilon}_n\equiv \epsilon_n/E_F$,
 and $\Psi_n$, the 
transpose of
\begin{equation}
\Psi_n^T =
(u^{\uparrow}_{n1},\ldots,u^{\uparrow}_{nN},u^{\downarrow}_{n1},\ldots,u^{\downarrow}_{nN},
v^{\uparrow}_{n1},\ldots,v^{\uparrow}_{nN},v^{\downarrow}_{n1},\ldots,v^{\downarrow}_{nN}), 
\end{equation}
contains the
expansion coefficients associated with the set of orthonormal
basis functions. We write
$u^{\alpha}_n(z)=\sqrt{{2}/{d}}\sum_{q=1}^N u^{\alpha}_{n q}\sin(q \pi z/d)$, and
$v^{\alpha}_n(z)=\sqrt{{2}/{d}}\sum_{q=1}^N v^{\alpha}_{n q}
\sin(q\pi z/d),$ for $\alpha=\uparrow,\downarrow$. 
The necessary matrix elements analogous to Eqn.~(\ref{bogo}) for different $\pi$ junction geometries 
and for strictly collinear magnetization orientations have been calculated in 
previous work.\cite{hv69,hv70} The situation is more complicated for the case
of a FSF trilayer 
considered here,  with the magnetization angles of the two F layers
forming an angle $\alpha$. 
The matrix elements 
are then written as,
\begin{subequations}
\begin{align}
(H_{0})_{m n}&=\left[\left(\frac{m\pi}{k_{F} d}\right)^2 +
\frac{\varepsilon_{\perp}}{E_F} -1\right]\delta_{m n}+Z_B[{\cal U}_{m-n}(d_{F1})+
{\cal U}_{m-n}(d_{F1}+d_S)-{\cal U}_{m+n}(d_{F1}) \nonumber \\ 
&-{\cal U}_{m+n}(d_{F1}+d_S)], \\
(H_z)_{m n}&=\frac{h_0}{E_F}\cos(\alpha/2)\bigl[ {\cal K}_{m-n}(d_{F1})-{\cal K}_{m+n}(d_{F1})
+{\cal K}_{m+n}(d_{F1}+d_S) \nonumber \\
&-{\cal K}_{m-n}(d_{F1}+d_S)\bigr],\quad  m\neq n, \\
&=\frac{h_0}{E_F}\cos(\alpha/2) \left[\frac{d_{F1}+d_{F2}}{d}
+{\cal K}_{2m}(d_{F1}+d_S)-{\cal K}_{2m}(d_{F1})\right], \quad m=n, \\
(H_x)_{m n}&=\frac{h_0}{E_F} \sin(\alpha/2)\bigl[ {\cal K}_{m+n}(d_{F1})-{\cal K}_{m-n}(d_{F1})+
{\cal K}_{m+n}(d_{F1}+d_S) \nonumber \\
&-{\cal K}_{m-n}(d_{F1}+d_S)\bigr],\quad  m\neq n, \\
&=\frac{h_0}{E_F} \sin(\alpha/2)\left[\frac{d_{F2}-d_{F1}}{d}+{\cal K}_{2m}(d_{F1}+d_S)+
{\cal K}_{2m}(d_{F1})\right], m=n, \\
(D)_{m n}&=\frac{2}{E_F d}\int_{d_{F1}}^{d_{F1}+d_S} \hspace{-.15cm} dy \sin\left[\frac{m \pi y}{d}\right] 
\Delta(y) \sin\left[\frac{n \pi y}{d}\right],
\end{align}
\end{subequations}
\end{widetext}
where $Z_B\equiv 2 H_B/(k_F d)$ is a convenient dimensionless
measure of interfacial scattering. We have also defined:
\begin{align}
{\cal K}_{n}(y)&\equiv \frac{\sin\left(\frac{n \pi y}{d}\right)}{n\pi}, \quad
{\cal U}_n(y)\equiv \cos\left(\frac{ n \pi y}{d}\right).
\end{align}

We now consider the appropriate quantities that  characterize the induced
triplet correlations. To do this, we
define the following 
{\it triplet} pair  amplitude functions in terms of the field operators,
\begin{subequations}
\label{pa}
\begin{align}
{\bf f_0}({\bf r},t) =& \frac{1}{2}[\langle \psi_{\uparrow}({\bf r},t) \psi_{\downarrow} 
({\bf r},0)\rangle+
\langle \psi_{\downarrow}({\bf r},t) \psi_{\uparrow} ({\bf r},0)\rangle],\\
{\bf f_1}({\bf r},t) =& \frac{1}{2}[\langle \psi_{\uparrow}({\bf r},t) \psi_{\uparrow} 
({\bf r},0)\rangle -\langle \psi_{\downarrow}({\bf r},t) \psi_{\downarrow} 
({\bf r},0)\rangle].
\end{align}
\end{subequations}
We will later demonstrate that
these amplitudes vanish at $t=0$, as required by the Pauli  principle. 

To make use of these expressions, it is most
convenient to use the Heisenberg
 picture. Thus we write 
$\psi_\varsigma$ in the Heisenberg representation:
\begin{equation}
\psi_\varsigma(t)=e^{(i{\cal H}_{\rm eff}t)} \psi_\varsigma e^{(-i{\cal H}_{\rm
eff}t)}.
\end{equation}
To put this in terms of the quasiparticle amplitudes, we apply
Eqns.~(\ref{bv}) and the transformation 
Eqns.~(\ref{comm2}). 
We can then immediately write down the Heisenberg
equations of motion for the $\gamma$'s as
\begin{equation}
i \frac{\partial \gamma_n}{\partial t}=[\gamma_n,{\cal H}_{\rm eff}]
\end{equation}
and
\begin{equation}
i \frac{\partial \gamma_n^\dagger}{\partial t}=
[\gamma_n^\dagger,{\cal H}_{\rm eff}].
\end{equation}
These equations of motion, given Eqns.~(\ref{comm2}), 
have the solutions  $\gamma_n(t)=\gamma_n e^{-i\epsilon_nt}$
and $\gamma_n^\dagger(t)=\gamma_n^\dagger e^{i\epsilon_nt}$.  
When we substitute these results into the above equations for  
${\bf f_0}$ and ${\bf f_1}$, taking into account Eqns.~(\ref{bv}) we
obtain:
\begin{subequations}
\label{pa3}
\begin{align}
{\bf f_0}(y,t) =& \frac{1}{2}\sum_n[u_{n\uparrow}(y) v_{n\downarrow}(y)-u_{n\downarrow}(y) 
v_{n \uparrow}(y)]
\zeta_n(t), \label{f0} \\
{\bf f_1}(y,t) =& -\frac{1}{2}\sum_n [u_{n\uparrow}(y) v_{n\uparrow}(y)+u_{n\downarrow}(y) v_{n\downarrow}(y)] 
\zeta_n(t),
\end{align}
\end{subequations}
where $\zeta_n(t)\equiv \cos(\epsilon_n t)-i\sin(\epsilon_n t)\tanh(\epsilon_n/2T)$.
The spatial dependence of 
the complex quantities ${\bf f_0}(y,t)$  and ${\bf f_1}(y,t)$
is, in our geometry, on the $y$ coordinate only. 
They vanish identically 
at $t=0$.

We will focus on in our study of the induced triplet correlations on the time
dependent quantities ${\bf f_0}(y,t)$  and ${\bf f_1}(y,t)$. 
Their existence at $t>0$ is allowed 
by the Pauli principle. It is also important to sort out when it is allowed
by the spin symmetries: when the axis of quantization
of the Cooper pairs is the only  axis of quantization in the system (i.e.,
when $\alpha=0$) then it is not hard to see that the total spin operator
${\bf S}$ of the Cooper pairs does not
commute with the Hamiltonian. This is best
seen directly from the matrix expression on the left side of Eqn.~(\ref{bogo}).
On the other hand,  $S_z$ and the Hamiltonian do commute in this case.
However, when $\alpha$ (and therefore $h_x$) is nonzero, then no
component of  ${\bf S}$ commutes with the effective Hamiltonian. From
this spin symmetry argument 
it follows that the 
induced amplitude
${\bf f_1}(y,t)$ may exist (at finite times) only at nonzero $\alpha$, 
while ${\bf f_0}(y,t)$ is
allowed ant any $\alpha$. For $\alpha=\pi$, when the
magnetizations are antiparallel and along the $x$ axis, no triplet amplitudes
with nonzero component along that axis can exist. The  matrix 
$\widehat{U}(\alpha)$
in Eq.~(\ref{rot}) can be used to verify this
by performing the corresponding spin
rotations.   
That the existence of certain quantities is consistent with
all symmetry properties does not mean that these quantities will indeed
be nonvanishing, and it certainly tells us nothing about the possible
range and behavior in space and time of these amplitudes. To determine
this requires detailed calculations.     

Also of considerable interest in F/S structures is the reverse proximity
effect: the leakage of magnetism out of the magnets and into
the superconductor. This can be characterized
by the local magnetization ${\bf m}(y)$. 
It is defined as,
\begin{equation}
{\bf m} =-\mu_B \langle \sum_\sigma \psi_\sigma^\dagger {\bsigma} \psi_\sigma \rangle,
\end{equation}
where $\mu_B$ is the Bohr magneton. The vector
${\bf m}$ has two components in the FSF geometry discussed. Both
components depend on $y$. They are:
\begin{widetext}
\begin{equation}
m_z(y)=-\mu_B\sum_n \left\lbrace \left[|u_{n\uparrow}(y)|^2 - |u_{n\downarrow}(y)|^2 \right]f_n 
+  \left[|v_{n\uparrow}(y)|^2 - |v_{n\downarrow}(y)|^2 \right](1-f_n) \right\rbrace,
\label{mz}
\end{equation}
and
\begin{equation}
m_x(y)=-2 \mu_B\sum_n \left\lbrace \left[u_{n\uparrow}(y)u_{n\downarrow}(y) \right]f_n 
+  \left[v_{n\uparrow}(y)v_{n\downarrow}(y) \right](1-f_n) \right\rbrace.
\label{mx}
\end{equation}
It is convenient to 
normalize these components to $-\mu_B (N_\uparrow + N_\downarrow)$, where $N_\uparrow = 
k_F^3 (1+I)^{3/2} /(6\pi^2)$, and $N_\downarrow = 
k_F^3 (1-I)^{3/2} /(6\pi^2)$. 
%

The proximity effects can also be examined through the local DOS, $N(y,\epsilon)$, given by,
\begin{equation}
N(y,\epsilon)=-\sum_n\lbrace[u_{n\uparrow}^2(y)+u_{n\downarrow}^2(y)]f^\prime(\epsilon-\epsilon_n)
+ [v_{n\uparrow}^2(y)+v_{n\downarrow}^2(y)]f^\prime(\epsilon+\epsilon_n)\rbrace,
\label{ldos}
\end{equation}
\end{widetext}
where $f^\prime=\partial f/\partial \epsilon$. We will be concerned mainly with the
DOS normalized to the DOS of a bulk (unpolarized) normal metal, $D_N(0)=k_F^3/(2\pi^2 E_F)$.

\section{Results}
\label{results}

\begin{figure}
\includegraphics[width=3.25in]{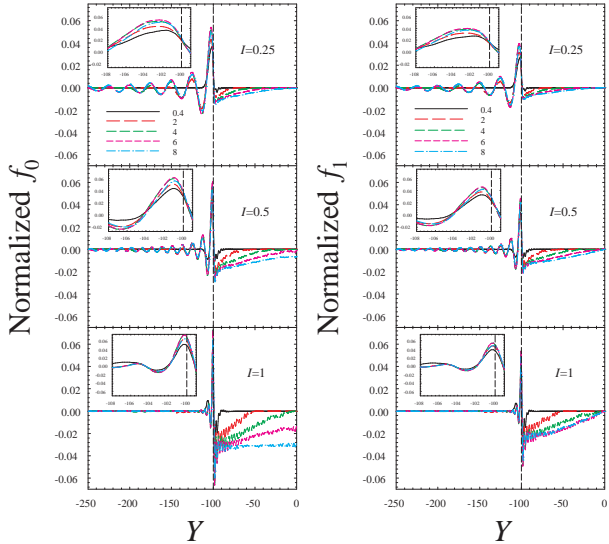}
\caption{The real parts, $f_0$ and $f_1$, of
the triplet pair amplitudes ${\bf f_0}$ and ${\bf f_1}$ (Eqns.~(\ref{pa3})), 
plotted as
a function of position (in terms
of $Y \equiv k_F y$) for three values of $I$
at different times $\tau \equiv \omega_D t$ indicated
in the legends of the top panels. These quantities are normalized to
the value of the singlet pairing amplitude in a bulk
S material. The main plots
show  
half of the S region (right side of the vertical dashed line), 
and part of the 
${\rm F_1}$ region. The insets are blow-ups of the region near
the interface. The angle $\alpha$ is zero in the left
panel and $\pi/2$ in the right panel.}
\label{fig2} 
\end{figure}

In this section we present our results, obtained self consistently
as explained above and in previous\cite{hv69,hv70,hbv}
work. We have assumed a coherence length $k_{F}\xi_0=100$.
We will choose a geometry in which the layers are
relatively thick: $k_F d_s\equiv D_S= 200$ (that is, two coherence
lengths) and $D_{F_{1}}=D_{F_{2}}=250$. These values ensure that the sample will 
be overall superconducting at temperatures up to about $1/3$ of 
the transition temperature $T_c$, 
of a pure S bulk sample. This was not the case for the
smaller values used in Ref.~\onlinecite{hbv} where the condensation
energy was quite small (see e.g. figure 6 in Ref.~\onlinecite{hv70}).
This allows us to study the temperature dependence of the quantities
involved over a broad range.
The most important parameters are the angle $\alpha$ and the  magnet
strength $I$. We will vary $\alpha$ in its full range between 0 and $\pi$
and give results for the values of $I$ of 0.25, 0.5, and unity. No triplet 
amplitudes arise at $I=0$ (when magnetism is absent). In the results presented
we have $Z_B=0$, when proximity effects are in general maximized.

In Fig.~\ref{fig2} we present  comprehensive results for  the real parts
of ${\bf f_0}(y,t)$ and ${\bf f_1}(y,t)$, which we denote simply
as $f_0(y,t)$ and $f_1(y,t)$ respectively. These are plotted in terms of the
dimensionless variable $Y \equiv k_F y$. 
The amplitudes
are normalized to the value of the usual singlet amplitude
in a pure bulk S sample. The temperature
is set to zero in this figure. In the main plots, half of
the S region and a portion (three fifths) of the left ($F_1$) region are included.
The corresponding portion on the $F_2$ side can be inferred from the geometry
and symmetry considerations.
Results are plotted at three values of $I$  and at a number
of finite times $\tau \equiv \omega_D t$ between 0.4 and 8 as indicated
in the legends. We have verified that at $t=0$ the computed triplet amplitudes
vanish identically, in agreement with the Pauli principle. This is true,
however, only when the calculation is performed to
self-consistency: {\it non-self consistent results invariably violate the
Pauli principle} near the interface. The results for $f_0$ are given at 
an angle $\alpha=0$ while those for $f_1$ are at $\alpha=\pi/2$.
At $\alpha=0$, $f_1$ vanishes identically since the $z$
component of the total spin is
then a  good quantum number. At $\alpha=\pi/2$, and short time scales,
the spatial dependences of the
two triplet components coincide, albeit with different signs 
in the two magnet regions,
due to the magnetization vectors having equal projections on the 
$x$ and $z$ axes.
At longer times, when the triplet amplitudes 
extend throughout the S layer and couple the two magnets, 
$f_0$  and $f_1$  deviate from one another.
The insets in each panel amplify 
and clarify the region
near the interfaces.

\begin{figure}
\includegraphics[width=3.25in] {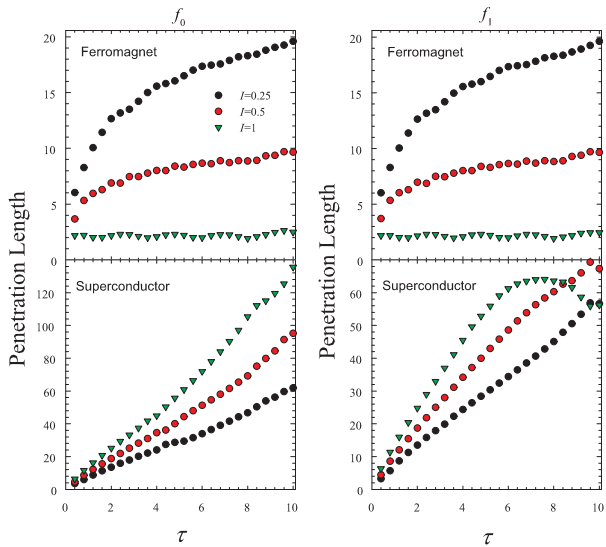}
\caption{Penetration depths for the triplet
amplitudes (see Eq.~(\ref{pl})), plotted as a function of $\tau$,
as calculated from $f_0$ and $f_1$
in both the S and F regions, for the values of $I$ indicated
and the same angles as in Fig.~\ref{fig2}.}
\label{fig3} 
\end{figure}

On the F side, both amplitudes peak very near the interface and then decay
in an oscillatory manner, reminiscent of the behavior of the usual
pair amplitude. Although the height of the first peak does not
depend strongly on $I$, the subsequent decay in the F material is faster
for larger values of $I$. This can be attributed to a decreased overall
proximity effect: here we have assumed
that at $I=0$ there would be no mismatch between the Fermi surface
wavevectors of the two materials, implying that as $I$
increases the mismatch between
either the up or the down Fermi wavevectors $k_\uparrow$
and $k_\downarrow$, on the F side, and that in the
S side increases. The location of this first peak depends very
clearly on $I$, its distance to the interface decreasing as 
approximately $1/I$ consistent with the general rule
that the oscillatory spatial dependences on the F side are
determined  by the inverse of $k_\uparrow - k_\downarrow$.
The height of the first peak depends strongly on  
time and is maximum at times $\tau$ of about $2 \pi$. It is quite
obvious that at intermediate values of $I$ the penetration of the triplet
correlation into the F material is rather long ranged.

\begin{figure}
\includegraphics[width=3.25in] {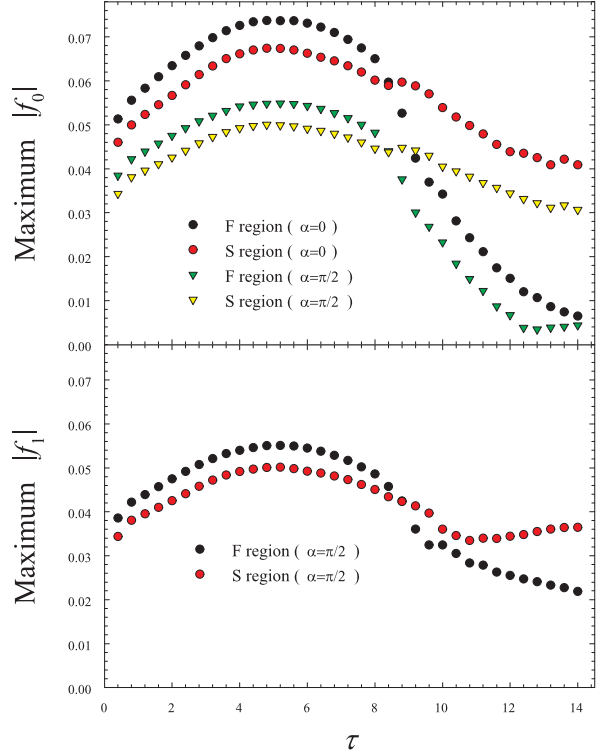}
\caption{Maximum absolute values of $f_0$ and $f_1$ (see text) as a function
of dimensionless time $\tau$ at $I=1$. In the top panel
we consider $f_0$ at both $\alpha=0$ and $\alpha=\pi/2$
while in the bottom panel we consider $f_1$ at $\alpha=\pi/2$. }
\label{fig4} 
\end{figure}

On the superconducting  side  the behavior is quite different: 
the triplet correlations penetrate into the S material over a distance
that rather quickly reaches two correlation lengths and then of course
saturates at the sample size, without signs of decaying in time
at these length scales. Furthermore
this effect now increases sharply with $I$ and is maximal in the half
metallic case. Thus, the magnets act as sources, so to speak, of triplet
correlations that enter the S material and this effect is stronger when $I$ is
larger. 

\begin{figure}
\includegraphics[width=3.25in] {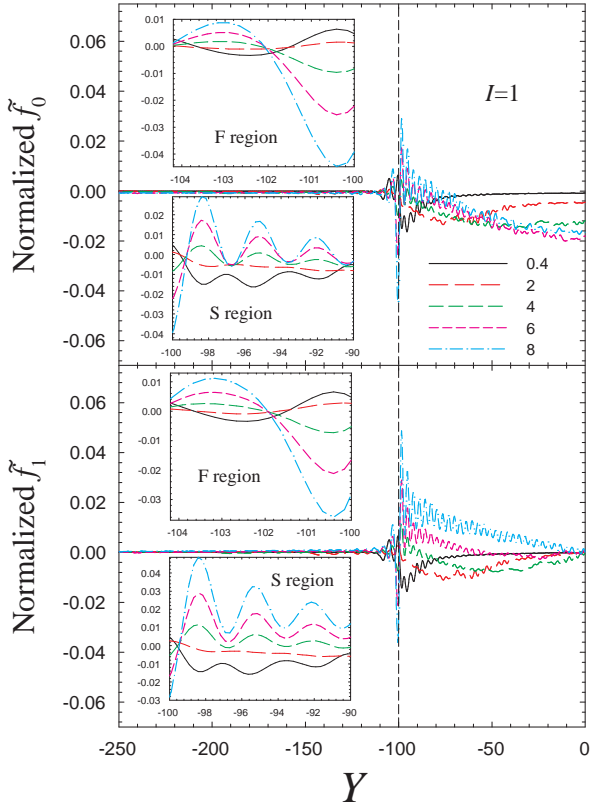}
\caption{Imaginary parts, $\tilde{f}_0(y,t)$ and $\tilde{f}_1(y,t)$
of the complex triplet amplitudes ${\bf f_0}(y,t)$ and ${\bf f_1}(y,t)$.
These quantities are normalized and plotted exactly as their
corresponding real parts are  in Fig.~\ref{fig2}, except that here
we consider only the $I=1$ case, and  both plots 
are  for $\alpha=\pi/2$. As in Fig.~\ref{fig2} the main
plot shows the behavior over an extended region. There are now two insets
to each main plot,
each showing the detailed behavior near the interface itself, on
either the S or F side.}
\label{fig5} 
\end{figure}

It is instructive to extract characteristic penetration lengths $\ell_i$ 
from the
above data using the definition,
\begin{align}
\label{pl}
\ell_i = \frac{\int dy |f_i (y,t)|}{\max|f_i(y,t)|}, \quad i=0,1,
\end{align}
where the integration is either over the S or the F region.
In Fig.~\ref{fig3}, the
top two panels show the penetration lengths for the F material, at 
three values of $I$ and the same values of $\alpha$ as for Fig.~\ref{fig2}.
The results are very similar
whether they are  calculated from the results for $f_0$ or from
those for $f_1$. The penetration length at constant time decreases with
$I$ as already noted and shows signs of saturating with time at a value which
for $I=0.25$ approaches that of the superconducting coherence length. On
the S side (bottom panels) the situation is very different:  the
results for $f_0$ and $f_1$ are now clearly dissimilar with the penetration
length for the former quantity being  (for cases 
shown here) the larger one. This arises from the geometry and 
magnetization projections of each F layer
on the $x$ axis, which are in opposite directions, forcing the 
triplet $f_1$ to possess a node at the center of the trilayer.
No such requirement exists for $f_0$, as it is spatially symmetric.
Except for the case of $f_1$ at $I=1$, we
see no sign of saturation. In fact, 
the maximum value of $\tau$ displayed here corresponds to
the case in which the entire intrinsically singlet superconductor layer, 
two coherence lengths 
thick and sandwiched between two 
magnets, is wholly pervaded by induced triplet correlations. 

It is also of interest to consider the variation of the spatial maximum values 
of $f_0(y,t)$ and $f_1(y,t)$ with time. 
In Fig.~\ref{fig4}
we show, for each time, the
largest value of these quantities, 
which typically is attained near the interface,  in either
the F or S regions. By ``maximum'' value we mean the maximum 
of $|f_0|$ and $|f_1|$,
not to be confused with the absolute value of the complex quantities
$|{\bf f_0}|$ or $|{\bf f_1}|$. In this figure, the magnets  are half 
metallic, $I=1$.
In the top panel, we plot the results for $f_0$ at both $\alpha=0$ and 
$\alpha=\pi/2$.
We see that at earlier times, the value $f_0(y,t)$ at its peak just inside the
F region (see Fig.~\ref{fig2}, bottom left
panel) exceeds the maximum value of this quantity in S. At longer times,
however, there is a crossover as the size of the peaks in F decreases rather
sharply,
as explained above, while the size of the amplitude in S decreases only slowly,
as the triplet correlations fill the S layer. It is apparent from careful
examination of the $I < 1$  panels in Fig.~\ref{fig2}, that this crossover 
does not occur for smaller values of $I$ except possibly on a time scale
much longer than that considered here. In the bottom panel, we
present  a  similar study  of $f_1$, this time of course only at $\alpha=\pi/2$
since this quantity vanishes identically for collinear magnetizations. The results are
clearly very similar except that the results in the F region appear to saturate
and do not decrease at long times. 
This is consistent with the earlier discussion where we saw that 
$|f_0|$ and $|f_1|$
overlap at $\alpha=\pi/2$, except at sufficiently long times.
In all cases the maximum value of the
quantity plotted crests near $\tau=6$ in agreement with  previous remarks.

\begin{figure}
\includegraphics [width=3.25in] {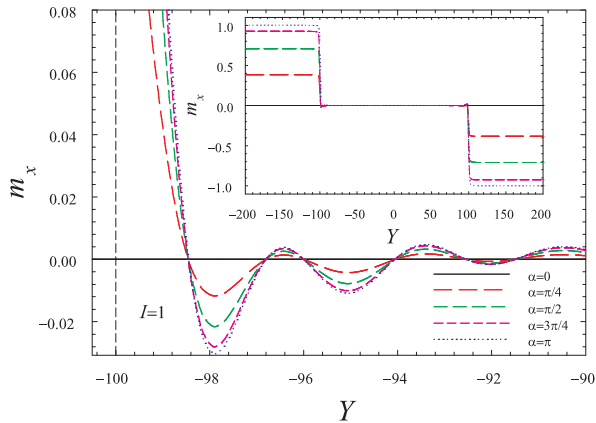}
\caption{Magnetic moment component $m_x$ (see
Eq.~(\ref{mx})), normalized as explained
in the text, plotted vs position (at $I=1$)
for several values of $\alpha$. The main
plot shows the behavior near the interface (vertical dashed line), 
while
the inset covers the whole sample. }
\label{fig6} 
\end{figure}

All of the above results have been given in terms of the real parts, $f_0$
and $f_1$ of the complex amplitudes ${\bf f_0}$ and ${\bf f_1}$. The behavior
of the corresponding imaginary parts is qualitatively very similar and thus we
will present only two examples, in Fig.~\ref{fig5}. We denote these
imaginary parts by $\tilde{f}_0(y,t)$ and $\tilde{f}_1(y,t)$ respectively.
In Fig.~\ref{fig5} we consider only the case $I=1$ (compare with Fig.~\ref{fig2})
and $\alpha=\pi/2$.  As in Fig.~\ref{fig2} the main plots include
an extended region near one of the interfaces and the insets  are close views
of the interface itself, in this case one of the insets shows a more 
detailed view of the S side. On the F side, the behavior is reminiscent
to that of the real parts, except that the very prominent peak seen in
the real parts right at the interface is absent for the imaginary parts.
On the S side, the sign is now initially negative and it changes to positive
at $\tau$ of order unity. No such change was observed for the real parts.
At longer times, 
the imaginary part of the triplet correlations also eventually penetrates
several correlation lengths into the S sample, just as  the real part does. 

\begin{figure}
\includegraphics [width=3.25in] {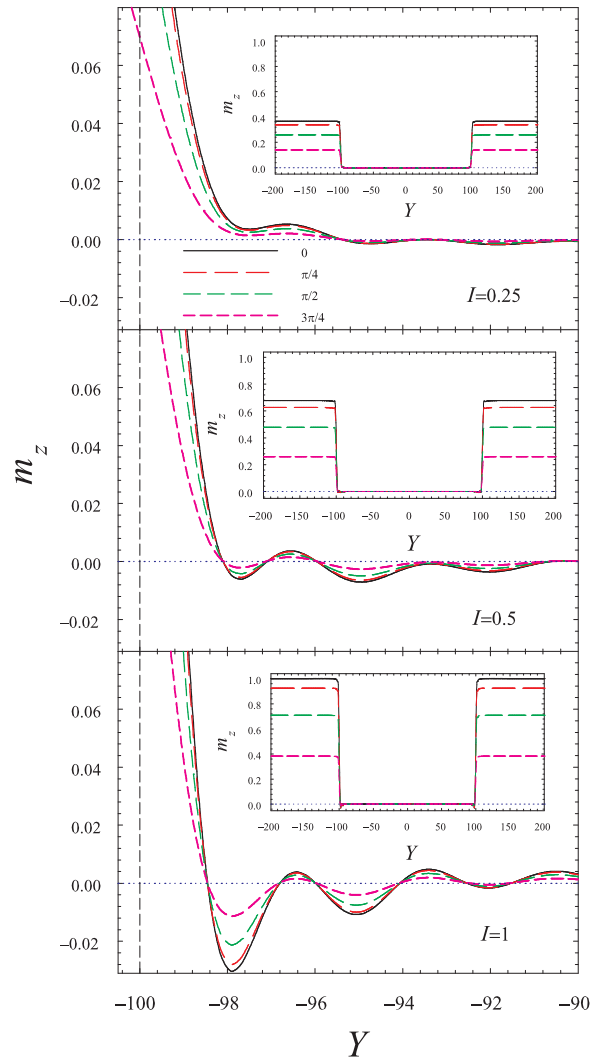}
\caption{Normalized (see text) magnetic moment component  $m_z$ (Eq.~(\ref{mz}))
plotted vs. dimensionless position
for three values of $I$. Again, the main
plot is the behavior near the interface, while the insets cover 
the whole sample. It is evident that $m_z$ points in the same direction for both magnets}
\label{fig7} 
\end{figure}

In the next two figures, we explore the reverse proximity effect (the
spreading of the magnetism into the S layer) as a function 
of $\alpha$. This is best done by considering separately the 
two components of the local magnetic
moment vector. First, in Fig.~\ref{fig6} we consider the $x$ component $m_x$
(see Eq.~(\ref{mx})) normalized 
to the absolute value of
its
bulk value in a pure F material. The results
in this figure are for 
half metallic magnets,  
($I=1$). In the main plot of the figure we 
display the value of $m_x$ in the region very near an interface
for several values of $\alpha$. Of course, $m_x$
vanishes at $\alpha=0$. At other values of $\alpha$ it is large
in the F material and it penetrates into S in an
oscillatory way that is quite
reminiscent of the corresponding  penetration of the superconducting
correlations into S. We see that the 
period of the spatial oscillations of
the magnetization 
is independent of the angle $\alpha$ between 
the two F layer magnetizations. Another discernible feature is that $m_x$ dampens out
over relatively short length scales, consistent with past work.\cite{hv69}
The inset shows the overall behavior of $m_x$
in the entire sample, demonstrating also the opposite signs between
in the two magnets in accordance
with Fig.~\ref{fig1}.

\begin{figure}
\includegraphics[width=3.25in]{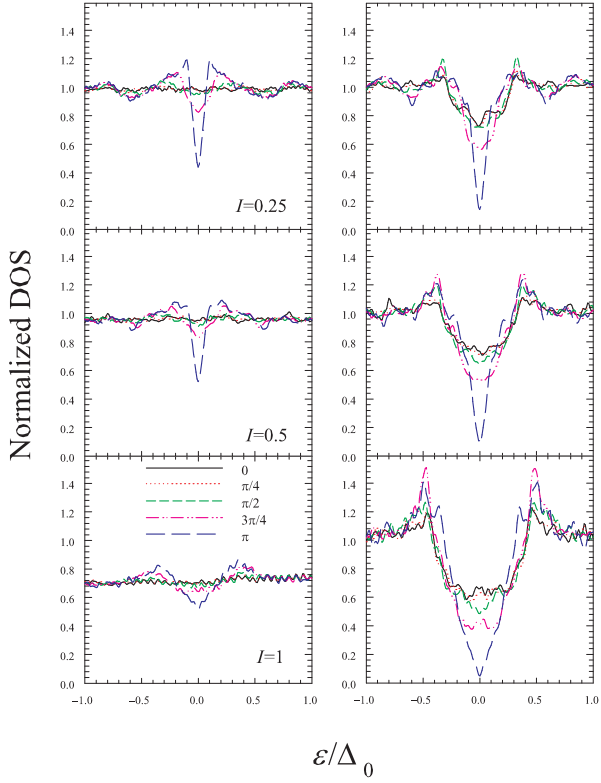}
\caption{Normalized local DOS from Eq.~(\ref{ldos}) integrated
over  (see text) over either the F region (left panels) or 
the S region (right panels) for three values of $I$ and several relative magnetization
orientations. The temperature is at $T=0.01 T_c$.}
\label{fig8} 
\end{figure}

Similarly, in Fig.~\ref{fig7} we display the 
$z$-component of $m_z$ (see Eq.~(\ref{mz})), normalized in the
same way as $m_x$, and for the same values of $\alpha$ but 
including now three different
values of $I$. Again, the main plots display the behavior
near the interface while the insets are for the entire sample. One can
see here that the reverse proximity effect is very weak at small $I$ and
largest in the half metallic case. The magnetic moment
oscillates in the superconductor 
with a period that is independent of the direction
and magnitude of the mutual magnetization in the F layers.
The observed trends in $m_z$ hold also for the $m_x$
component.

\begin{figure}
\includegraphics[width=3.25in]{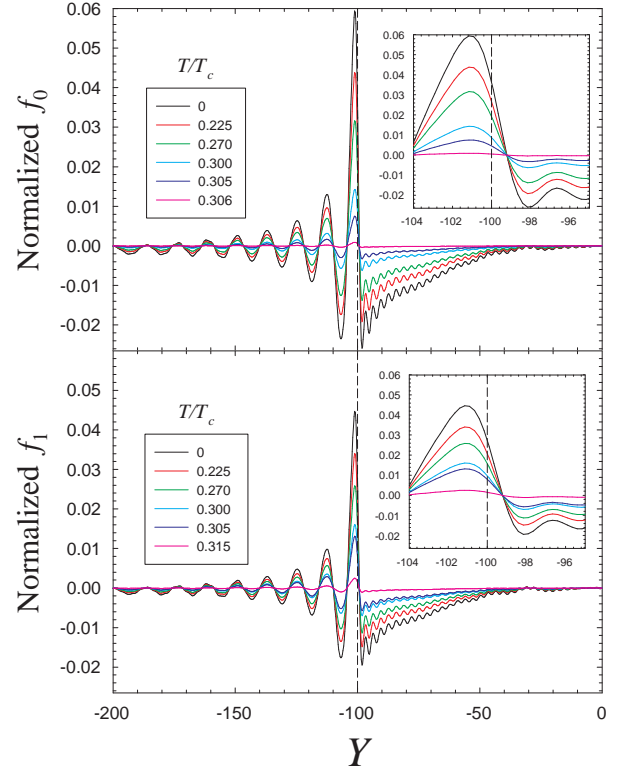}
\caption{Temperature 
and position dependence of $f_0$ and $f_1$ at $I=0.5$ and
$\tau=4$. In the top panel $\alpha=0$ and in the bottom
panel $\alpha=\pi/2$. The
values of $T/T_c$, where $T_c$ is the transition temperature
of bulk S, are indicated. The main plots show a rather wide
region near the interface (vertical dashed line) and the insets
focus on the region very near the interface.}
\label{fig9} 
\end{figure}

We next study the energy dependence of the single particle quasiparticle
spectrum by considering the local density of states (local DOS),
$N(y,\epsilon)$, as defined in Eqn.~(\ref{ldos}). 
In Fig.~\ref{fig8} we
consider the local DOS, 
integrated either over either the entire S or the entire F region, and
normalized to its bulk value on a sample of the S material in its
normal (non-superconducting) state.  The results are displayed
for three values of $I$ and several values of $\alpha$. The results 
reflect and confirm  what we already have found out from analyzing
the triplet amplitudes. On the F side, the proximity effect increases the correlations
somewhat with $\alpha$, as a
reduction in quasiparticle states emerges for low energies due to increased
correlations as the relative magnetizations become increasingly
antiparallel. As $I$ increases, the proximity effects
weaken. Indeed,
at $I=1$ the DOS is nearly flat except at the larger values of $\alpha$,
suggesting that in the absence of a down spin band there
may be a contribution from a possible triplet presence.
On the superconducting side, the situation is somewhat different: the
results never approach a limit where the local DOS would look similar to
that of a bulk superconductor. Even at $I=1$, there is never a gap and
even when that situation is approached at larger $\alpha$, when the
magnetizations in the two F layers are antiparallel, the shape of the
DOS curve 
does not resemble the signature bulk superconductor result. The
profound difference between the parallel ($\alpha=0$) and antiparallel
($\alpha=\pi$) cases is consistent in every respect with that 
previously\cite{hv72} found. 

\begin{figure}
\centering
\includegraphics [width=3.25in] {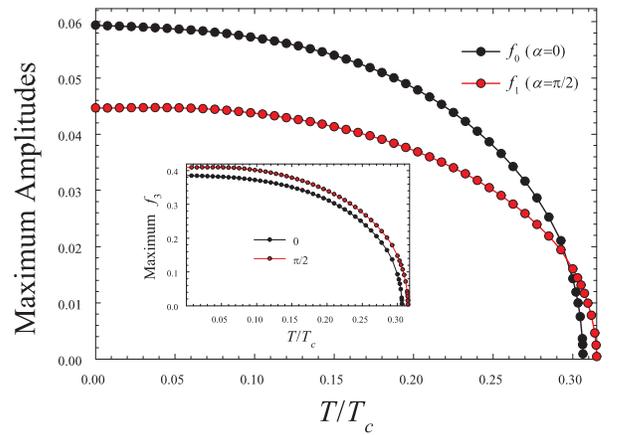}
\caption{The maximum values of the real parts of
the triplet amplitudes at $I=0.5$ and $\tau=4$, as a function
of temperature and the same values of $\alpha$ as
in Fig.~\ref{fig9}. The inset shows the corresponding peak values
of the ordinary singlet amplitude, $f_3\equiv\Delta(y)/g$ (Eq.~(\ref{del2})).}
\label{fig10} 
\end{figure} 

The above results were all obtained in the low temperature limit.
In the remaining figures we consider the temperature dependence.
In Fig.~\ref{fig9} we plot
directly the spatial behavior of $f_0$ and $f_1$ over a broad range of
temperatures.
As in previous figures for the triplet amplitudes,
the main plot shows the behavior over  a relatively extended region of the sample
and the insets magnify the region near 
interface. In both cases we have
taken $I=0.5$ and $\tau=4$ while $\alpha=0$ for $f_0$ and $\alpha=\pi/2$
for $f_1$. It can be inferred that at low T 
the temperature dependence of the triplet amplitudes
is weak, while as $T$ increases, the closer spacing in temperatures shown suggest that
the correlations become destroyed at a much more rapid rate.
In comparing the two panels, it is seen that for much of the temperature range, $|f_0|>|f_1|$,
but at higher temperatures ($T \gtrsim 0.3 T_c$), 
$f_1$ becomes the larger of the two.
The temperature dependence of the characteristic
penetration depths is weak. 
These observations are further exemplified
in Fig.~\ref{fig10}, where the peak values of $|f_0|$
and $|f_1|$ (also at $\tau=4$) are shown as a function of $T$.
These quantities are determined by calculating 
$\max \lbrace f_0(y,T)\rbrace$ and $\max \lbrace f_1(y,T)\rbrace$
throughout the structure for a given temperature. 
The inset depicts the corresponding peak values of the ordinary self-
consistent equal-time singlet amplitude
$f_3(y)\equiv\Delta(y)/g$,
 (see Eq.~(\ref{del2})). 
For both the triplet and singlet behavior, there is a strong dependence on $T$ as the temperature approaches the
system's transition temperature, which is about $0.32 T_c$ for our
system.
 Technically, the determination
of the self consistent amplitudes is more difficult at 
higher $T$, when the number of iterations is in principle much higher.
This increase in computational time can be reduced
by up to an order of magnitude by taking as the initial
spatial pair potential at a given $T$ the result for the previously obtained
next lower temperature times
a $T$ dependent factor
derived from the linearized Ginzburg-Landau theory\cite{bdg}.

\begin{figure}
\includegraphics[width=3.25in]{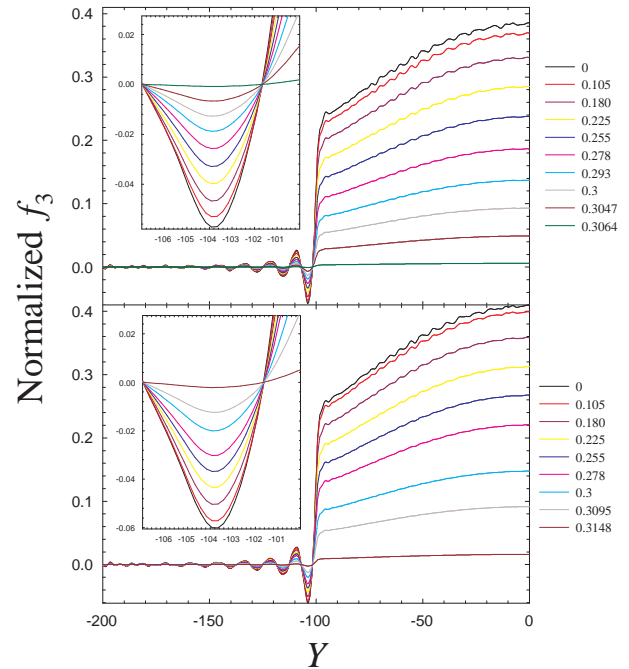}
\caption{Temperature and position dependence of the ordinary
singlet pair amplitude (or $\Delta(y)/g)$), 
normalized to its value in bulk S material, for moderate magnetic strength, $I=0.5$. 
Top panel: the magnetizations are parallel ($\alpha=0$).
Bottom panel: the magnetizations are perpendicular ($\alpha=\pi/2$).}
\label{fig11} 
\end{figure}
In accordance with what we have just seen, it is a natural extension to 
study how thermal effects
might destroy the spatial characteristics of
singlet correlations throughout the structure.
In
Fig.~\ref{fig11}, we therefore display $f_3$ 
as a function of $Y$, for several temperatures, and 
in which the relative magnetizations are collinear ($\alpha=0$)
and at right angles ($\alpha=\pi/2$).
Remarkably, one can see that the temperature dependence 
of the triplet components is somewhat weaker that that  of the standard
singlet amplitude. One can also clearly see that, as indicated above,
the penetration of the triplet amplitudes into the F material over a length
scale that is clearly much longer than that of the singlet amplitude. This
is again a strong indication that experimental tunneling results
indicating long ranged penetration effects in F/S structures are indeed
evidence for induced triplet correlations.

\section{Conclusions}
\label{conclusions} 
In this paper, we have presented a detailed study of induced 
time dependent (odd in time) triplet
pairing correlations in  clean 
planar FSF junctions consisting of an 
ordinary $s$-wave superconductor
sandwiched between two relatively thick ferromagnets whose magnetizations are misoriented
with respect to each other by an angle $\alpha$. 
Our microscopic formalism allowed us to investigate
cases involving strong magnets, as well as atomic scale phenomenon,
two things not possible in the widely used quasiclassical approaches.
We have obtained  results
as a function of $\alpha$, time,  the strength $I$ of the ferromagnets, and the
temperature. We have presented results for 
the spatial behavior of  the time-dependent
triplet pair amplitudes, $f_0$ and $f_1$, and for the corresponding 
penetration lengths extracted from them. We have found that these triplet
correlations are indeed induced via the proximity effect, that
they completely pervade even a superconductor several coherence lengths 
thick, and also substantially penetrate the ferromagnetic layers. These
results have clear implications for the experimental work in which
long range proximity effects in SF nanostructures have been reported,
effects that have been speculated to be due to the existence of some kind of 
triplet pairing. Our calculations, in which the time dependence is
studied from the Heisenberg picture, emphasize the need for full
self-consistency of the solutions, without which we find that the Pauli
principle is violated.

We have also considered 
the reverse proximity
effect, which is of particular interest in this case due to the presence
of two components of the magnetization, and we also have given results for
the experimentally  measurable local density of states,
which revealed clear subgap energy signatures as a function of $\alpha$ and $I$.
We have 
studied the temperature dependence of the triplet amplitudes (as well
as the ordinary singlet amplitude) and found that the temperature
dependence of the penetration depths associated with these 
triplet amplitudes is weak:
these lengths remain large all the way up to the vicinity of the transition temperature 
of the system.  
This bodes well for further experimental observations and verification
in these clean systems.

\acknowledgments
This project is
funded in part by the Office of Naval Research
(ONR) In-House Laboratory Independent Research (ILIR) Program and by a grant of HPC resources from 
the Arctic Region Supercomputing Center at the University of Alaska Fairbanks 
as part of the Department of Defense High Performance Computing Modernization 
Program.


\begin{thebibliography}{99}
\bibitem{sfhe} D.D. Osheroff, R.C. Richardson and D.M. Lee, \prl {bf 28}, 885
 (1972).

\bibitem{berez} V.L. Berezinskii, JETP Lett.~{\bf 20}, 287, (1974).


\bibitem{bulgac} Aurel Bulgac, Michael McNeil Forbes, and Achim Schwenk, \prl
{\bf 97}, 020402 (2006).
\bibitem{zutic} Igor \u{Z}uti\'{c}, Jaroslav Fabian, S. Das Sarma, \rmp {\bf 76}, 323
(2004).

\bibitem{gu} J. Y. Gu, 
C.-Y. You, J. S. Jiang, J. Pearson, Ya. B. Bazaliy, 
and S. D. Bader, \prl {\bf 89}, 267001 (2002).

\bibitem{moraru} I. C. Moraru, W. P. Pratt, N. O. Birge, \prl {\bf 96}, 037004 (2006).

\bibitem{bell} C. Bell, S. Tur\c{s}ucu, and J. Aarts, \prb {\bf 74}, 214520 (2006).

\bibitem{visani} C. Visani, 
V. Pe\~{n}a, J. Garcia-Barriocanal, D. Arias, Z. Sefrioui, C. Leon,  
J. Santamaria, N.M. Nemes, M. Garcia-Hernandez,  J.L. 
Martinez, S.G.E. te Velthuis and A. Hoffmann, 
\prb {\bf 75}, 054501 (2007).

\bibitem{hv72} K. Halterman and O.T. Valls, \prb {\bf 72}, 060514(R) (2005).

\bibitem{dg66} P. G. de Gennes, Phys. Lett. {\bf 23}, 10 (1966).

\bibitem{tagirov} L.R. Tagirov, \prl {\bf 83}, 2058 (1999).

\bibitem{buzdin99} A.I. Buzdin, A.V. Vdyayev, and N.V. Ryzhanova, 
Europhys. Lett. {\bf 48}, 686 (1999).

\bibitem{pena} V. Pe\~{n}a, Z. Sefrioui, D. Arias, C. Leon, J. Santamaria, M. Varela, S. J. Pennycook, J. L. Martinez, \prb {\bf 69}, 224502 (2004).

\bibitem{nelson} K. D. Nelson, Z. Q. Mao, Y. Maeno, Y. Liu, Science {\bf 306}, 1151 (2004).

\bibitem{kriv} V.N. Krivoruchko and V. Yu. Taernkov \prb {\bf 75}, 214508 (2007).

\bibitem{klapwijk06} R. S. Keizer, S. T. B. Goennenwein, T. M. Klapwijk, G. Miao, G. Xiao, A. Gupta, {\em Nature} {\bf 439}, 825, (2006).  

\bibitem{berg05} F.S. Bergeret,  A.F Volkov, and K.B. Efetov, \rmp {\bf 77}, 1321 (2005).

\bibitem{hbv} K. Halterman, P.H. Barsic, and O.T. Valls \prl {\bf 99}, 127002 (2007).  

\bibitem{berg01} F.S. Bergeret,  A.F Volkov, and K.B. Efetov, \prl {\bf 86}, 3140 (2001).

\bibitem{berg03} F. S. Bergeret, A. F. Volkov, and K. B. Efetov, \prb {\bf 68}, 064513 (2003).

\bibitem{eschrig05a} T. Champel and M. Eschrig, \prb {\bf 72}, 054523 (2005).

\bibitem{fominov07} Ya. V. Fominov, A. F. Volkov, and K. B. Efetov \prb {\bf 75}, 104509 (2007).

\bibitem{braude} V. Braude and Yu. V. Nazarov \prl {\bf 98}, 077003 (2007).

\bibitem{houzet07} M. Houzet and A.I. Buzdin \prb {\bf 76}, 060504(R) (2007).

\bibitem{fominov}Ya. V. Fominov, A. A. Golubov, and M. Yu. Kupriyanov, JETP Lett. {\bf 77}, 510 (2003).

\bibitem{eschrig05b} T. L\"{o}fwander, T. Champel, J. Durst, and M. Eschrig, \prl {\bf 95}, 187003 (2005).

\bibitem{buzdin05} A.I. Buzdin, \rmp {\bf 77}, 935 (2005).

\bibitem{asano} Y. Asano, Y. Sawa, Y. Tanaka, and A. Golubov, \prb {\bf 76}, 224525 (2007).

\bibitem{yokoyama} T. Yokoyama, Y. Tanaka, and A.A. Golubov, \prb {\bf 75}, 134510 (2007).

\bibitem{tanaka} Y. Tanaka, Y. Tanuma, and A. A. Golubov, \prb {\bf 76}, 054522 (2007).

\bibitem{linder} J. Linder and A. Sudb\o, \prb {\bf 75}, 134509 (2007).


\bibitem{bdg} P.G. de Gennes, {\it Superconductivity of Metals and Alloys},
(Addison-Wesley, Reading, MA, 1989). 
\bibitem{song}  J.~B. Ketterson and S.~N. Song, {\it Superconductivity},
(Cambridge  University Press, Cambridge, UK 1999), p. 286.
\bibitem{hv69} K. Halterman and O.T. Valls, \prb {\bf 69}, 014517 (2004).
\bibitem{hv70} K. Halterman and O.T. Valls, \prb {\bf 70}, 104516 (2004).
 
\end{thebibliography}
\end{document}